\documentclass[twocolumn]{aastex631}

\usepackage{amsmath, amssymb}

\newcommand{\Sp}{{\it Spitzer\/}}
\newcommand{\Ch}{{\it Chandra\/}}

\shorttitle{Multi-wavelength Sgr A* Observations. I. 2019 July 18}
\shortauthors{Michail et al. (2021c)}
\received{2021 July 19}
\revised{2021 October 1}
\accepted{2021 October 4}
\submitjournal{ApJ}
\graphicspath{{./figures}}
\begin{document}

\title{Multi-wavelength Observations of Sgr A*. I. 2019 July 18}

\author[0000-0003-3503-3446]{Joseph M. Michail}
\affiliation{Center for Interdisciplinary Exploration and Research in Astrophysics (CIERA) and Department of Physics and Astronomy, Northwestern University, 1800 Sherman Ave., Evanston, IL 60201, USA}
\correspondingauthor{Joseph M. Michail}
\email{michail@u.northwestern.edu}

\author[0000-0002-1737-0871]{Mark Wardle}
\affiliation{Research Centre for Astronomy, Astrophysics and Astrophotonics and Department of Physics and Astronomy, Macquarie University, Sydney, NSW 2109, Australia}

\author{Farhad Yusef-Zadeh}
\affiliation{Center for Interdisciplinary Exploration and Research in Astrophysics (CIERA) and Department of Physics and Astronomy, Northwestern University, 1800 Sherman Ave., Evanston, IL 60201, USA}

\author[0000-0002-1568-579X]{Devaky Kunneriath}
\affiliation{National Radio Astronomy Observatory, 520 Edgemont Road, Charlottesville, VA 22903, USA}



\begin{abstract}
We present and analyze ALMA submillimeter observations from a multi-wavelength campaign of Sgr A* during 2019 July 18. In addition to the submillimeter, we utilize concurrent mid-IR (\Sp) and X-ray (\Ch) observations. The submillimeter emission lags less than $\delta t\approx30$ minutes behind the mid-IR data. However, the entire submillimeter flare was not observed, raising the possibility that the time delay is a consequence of incomplete sampling of the light curve. The decay of the submillimeter emission is not consistent with synchrotron cooling. Therefore, we analyze these data adopting an adiabatically expanding synchrotron source that is initially optically thick or thin in the submillimeter, yielding time-delayed or synchronous flaring with the IR, respectively. The time-delayed model is consistent with a plasma blob of radius $0.8~R_{\text{S}}$ (Schwarzschild radius), electron power-law index $p=3.5$ ($N(E)\propto E^{-p}$), equipartition magnetic field of $B_{\text{eq}}\approx90$ Gauss, and expansion velocity $v_{\text{exp}}\approx0.004c$. The simultaneous emission is fit by a plasma blob of radius $2~R_{\text{S}}$, $p=2.5$, $B_{\text{eq}}\approx27$ Gauss, and $v_{\text{exp}}\approx0.014c$. Since the submillimeter time delay is not completely unambiguous, we cannot definitively conclude which model better represents the data. This observation presents the best evidence for a unified flaring mechanism between submillimeter and X-ray wavelengths and places significant constraints on the source size and magnetic field strength. We show that concurrent observations at lower frequencies would be able to determine if the flaring emission is initially optically thick or thin in the submillimeter.
\end{abstract}


\section{Introduction}
    Sagittarius A* (Sgr A*) is the $4.297\times10^{6}$ M$_\odot$ supermassive black hole located at a distance of $8.27$ kpc \citep{Gravity2019} in the center of the Milky Way. It is known to be a variable source at radio through X-ray wavelengths \citep[e.g.,][]{FYZ2006b, DoddsEden2009, Dexter2014, Hora2014} and is a prime target to understand how gas is captured, accreted, and/or ejected from low-luminosity supermassive black holes.
    
    At infrared (IR) and X-ray wavelengths, the emission from Sgr A* is optically thin synchrotron radiation \citep[e.g.,][]{Eckart2004, Eckart2006,Eckart2012, Ponti2017, Witzel2021}. At submillimeter (submm) and radio wavelengths, the variable emission's timescale is much shorter than that of the synchrotron cooling time. \citet{FYZ2006b} used an adiabatically expanding synchrotron plasma \citep{VDL1966} to describe the emission observed in this frequency range with much success. Since then, there has been a formidable effort to connect the radio/submm and IR/X-ray regimes with a model that can explain the variable emission across the electromagnetic spectrum. Multi-wavelength campaigns \citep[e.g.,][]{FYZ2008, Eckart2009, Mossoux2016, Ponti2017, Subroweit2020, Witzel2021} are critical to test this picture. Non-overlapping observations (due to the rise/set times at different observatories) and time-delayed emission at radio frequencies make this a difficult task. This has been somewhat alleviated by space-based IR instruments, like the Spitzer Space Telescope, in concert with submm observations, like ALMA \citep[e.g.,][]{Witzel2021}. Submm observations are key as they lay near the ``submillimeter bump," \citep{Zylka1992, Zylka1995} where Sgr A* is brightest. This bump begins in the millimeter (mm) and decays at wavelengths near the far-IR, where the emission becomes increasingly optically thin, and optical depth effects are unimportant \citep{Marrone2006}.
    
    \citet{Gravity2021} recently published an analysis from a multi-wavelength campaign on 2019 July 18 using IR and X-ray data. Here, we present simultaneous ALMA observations, which show a time delay relative to the mid-IR data, suggesting an optically thick and dynamically evolving source.
    
\section{Observations and Data Reduction}\label{sec:observations}
    The campaign included data from the Karl G. Jansky Very Large Array (VLA) and Giant Metrewave Radio Telescope (GMRT) on 2019 July 18; however, these light curves are not useful for this date. The VLA data at Q-band could not be calibrated by the VLA pipeline owing to bad weather and long cycle times. Extended structures in the GMRT data dominated Sgr A*, and identifying the intrinsic variability of Sgr A* is difficult. A more detailed accounts of these data will be given in J. Michail et al. (2021, in preparation).
    \subsection{Submillimeter: ALMA}
        The Atacama Large Millimeter/submillimeter Array (ALMA) data used were taken as part of a Cycle 6 Atacama Compact Array (ACA) observatory filler program (project code 2018.A.00050.T), which observed the Galactic Center in Band 7 on 2019 July 18 for about 7 hours. The dataset was calibrated and imaged by the ALMA pipeline (Pipeline-CASA54-P1-B, r42254, \texttt{CASA v5.4.0-70}, \citealp{McMullin2007}). The calibrated measurement set was obtained by restoring the calibration from the ALMA archive using \texttt{scriptForPI.py}. The data consist of 4 spectral windows of 2 GHz bandwidth and a spectral resolution of 1.953 MHz, with one spectral window centered on the CO transition at 346 GHz and three continuum spectral windows. J1337-1257 and J1924-2914 were used as bandpass and amplitude calibrators. J1700-2610 (J1700) and J1717-3342 were used as phase calibrators throughout the observation. J1733-3722 (J1733) was the phase calibrator during the final execution block. There were calibration issues in the gain amplitudes between the first three execution blocks; this was readily apparent in the light curve of J1700-2610, which showed that the flux varied by approximately $10\%$. To fix these offsets, we calculate average scale factors for the second and third execution blocks relative to the first using J1700-2610. These amplitude scale factors are applied to Sgr A* and J1700-2610. We chose the first execution block as our standard as its flux was most similar to J1700-2610's on the night of 2019 July 19. We do not complete the same scaling procedure for the last execution block as the phase calibrator is J1733-3722.
        %There was an offset in the fluxes ($\approx10\%$) between the first three executions, which used the same phase calibrator. To correct this offset, we scaled the fluxes of the second and third execution blocks to match the first execution block. 
        The observation was imported into \texttt{AIPS} for phase self-calibration. Then, \texttt{DFTPL} was used to extract the light curves from Sgr A* at a $60$ second binning time. The light curve for each spectral window is calculated independently using baselines greater than $20~\text{k}\lambda$. The ALMA light curves are shown in Figure \ref{fig:lcs} (left).
    
    \subsection{Mid-infrared: Spitzer}
        We obtained the binned \Sp\ light curves of Sgr A* for this day of observation (\Sp\ Sgr A* Collaboration, 2021, priv. comm.), which were originally published in \citet{Gravity2021}. Details of these observations are given in the aforementioned paper. Briefly, the IRAC instrument \citep{Fazio2004} on the Spitzer Space Telescope \citep{Werner2004} observed Sgr A* as part of \Sp\ program 14026 \citep{Fazio2018} at 4.5 \micron. The total on-source time for this observation was approximately $16$ hours. \citet{Hora2014} and \citet{Witzel2018} give further details on the observing mode and data reduction. The light curve time stamps were originally in the barycentric UT system. We converted these to UTC using an online tool \citep{Eastman2010}\footnote{The online tool is located here: \url{https://astroutils.astronomy.osu.edu/time/bjd2utc.html}}. The fluxes shown in Figure \ref{fig:lcs} (center) are relative to a period where Sgr A* is semi-quiescent and are not absolute \citep{Hora2014}. These data are de-reddened using the extinction value in \citet{Fritz2011} for the Band 2 IRAC instrument ($\text{A}_{4.5~\mu\text{m}} = 1.06$ mag). This yields a correction factor $f_{4.5~\mu\text{m}} = 2.512^{1.06} = 2.65$. The de-reddened \Sp\ light curve for 2019 July 18 is shown in Figure \ref{fig:lcs} (center).
        
    \subsection{X-ray: Chandra}
        Three days of simultaneous observations of Sgr A* were completed with the Chandra X-Ray Observatory. The corresponding observation ID for 2019 July 18 is 22230 (PI: Fazio) and was first published in \citet{Gravity2021}. This observation lasted 52 ks using \Ch's FAINT mode. Sgr A* was centered on the ACIS-S3 chip using the 1/8 subarray mode to reduce the possibility of pileup events from bright Galactic Center sources. The data are reprocessed with \texttt{CHANDRA\_REPRO} using \texttt{CIAO} v4.13 and \texttt{CALDB} v4.9.4. The WCS solutions are corrected using the \texttt{wcs\_match} and \texttt{wcs\_update} scripts utilizing 3 X-ray bright sources near Sgr A* from the Chandra Source Catalogue \citep[][CSC]{Evans2010} in \texttt{DS9}. One of these three sources is the nearby magnetar (SGR J1745-2900) that was readily apparent. The norm of the WCS correction is about 2 pixels ($1$\arcsec). %\sout{Finally, the \texttt{axbary} script is used to apply barycentric time corrections to the time stream data.
        
        To obtain background-subtracted light curves of Sgr A*, we follow \citet[and references therein]{Capellupo2017}. A 1\farcs25 radius aperture is centered at the J2000.0 radio location of Sgr A* ($17^\text{h} 45^\text{m} 40\fs04, -29^\circ00\arcmin28\farcs18$) to minimize contamination from the nearby magnetar. We only extract events within the $2-8$ keV range to minimize photons from the diffuse X-ray background \citep[e.g.,][]{Neilsen2013}. An aperture of radius $10\arcsec$ is centered on $17^\text{h}45^\text{m}40\fs30, -28\degr59\arcmin52\farcs66$ as the background region. The CSC is used to check for known sources within this area. No obvious point or extended sources are in this region which would bias our results. A 300 second binning time is adopted following previous light curve analyses of this source.

\begin{figure*}
        \centering
        \includegraphics[width=0.3\textwidth]{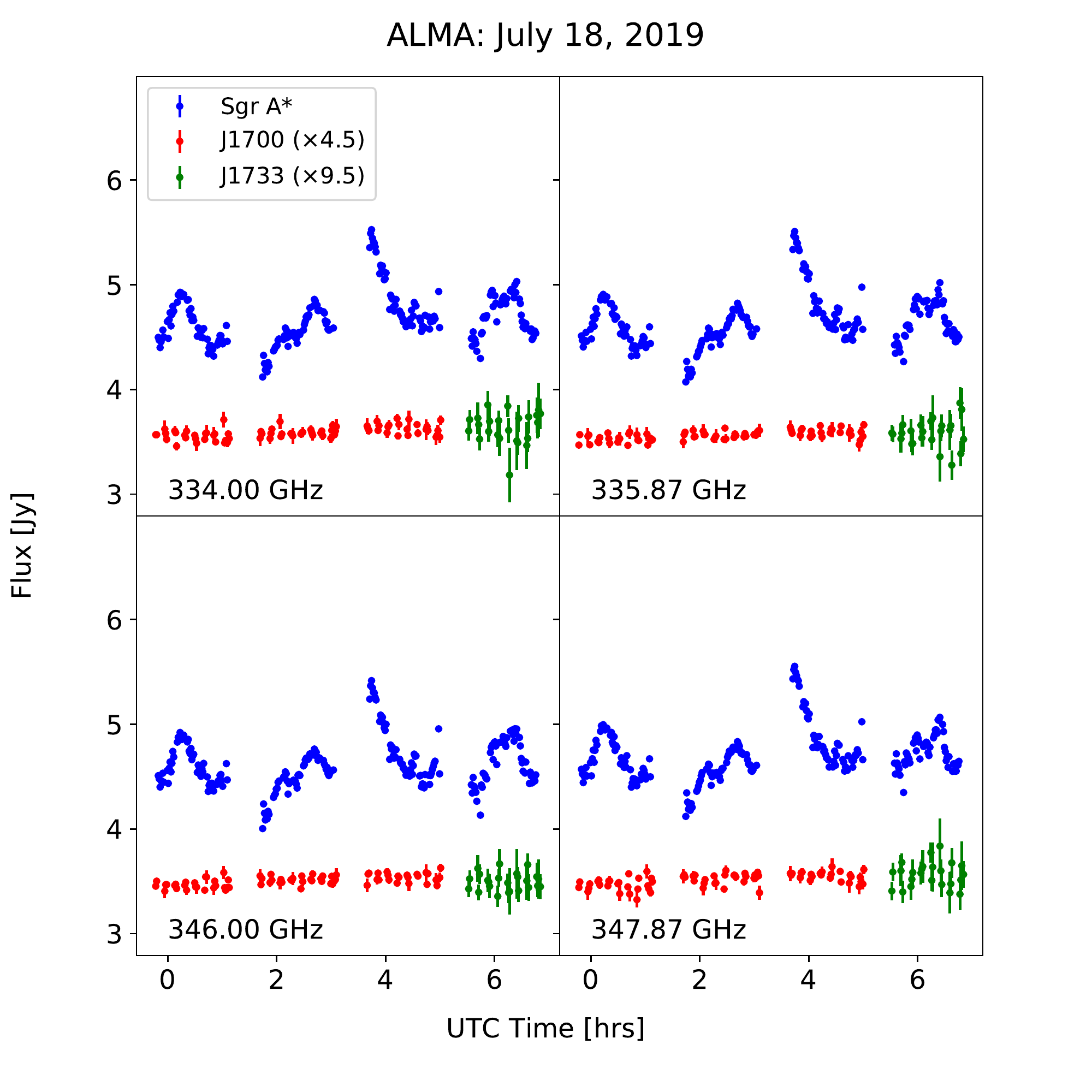}
        \includegraphics[width=0.3\textwidth]{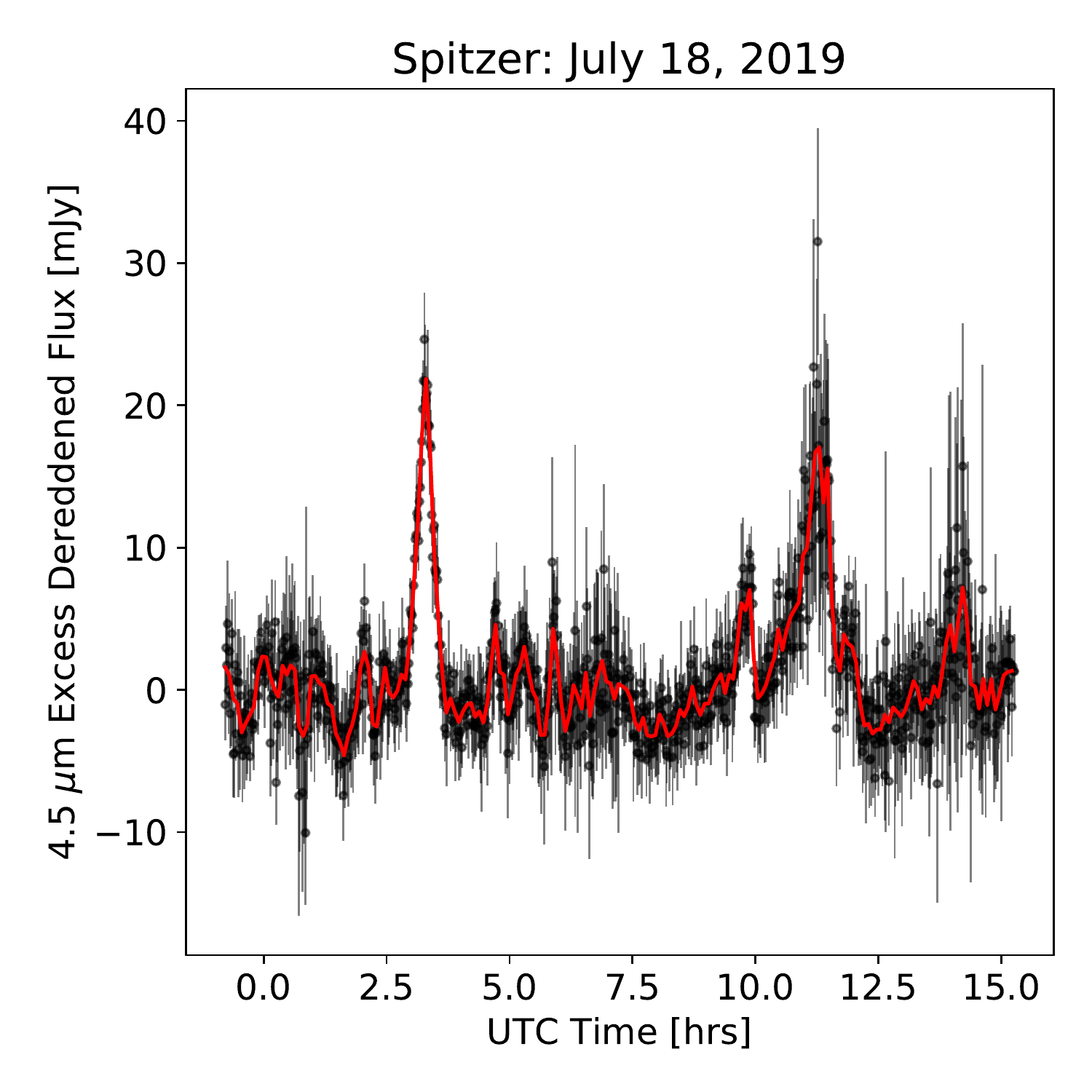}
        \includegraphics[width=0.3\textwidth]{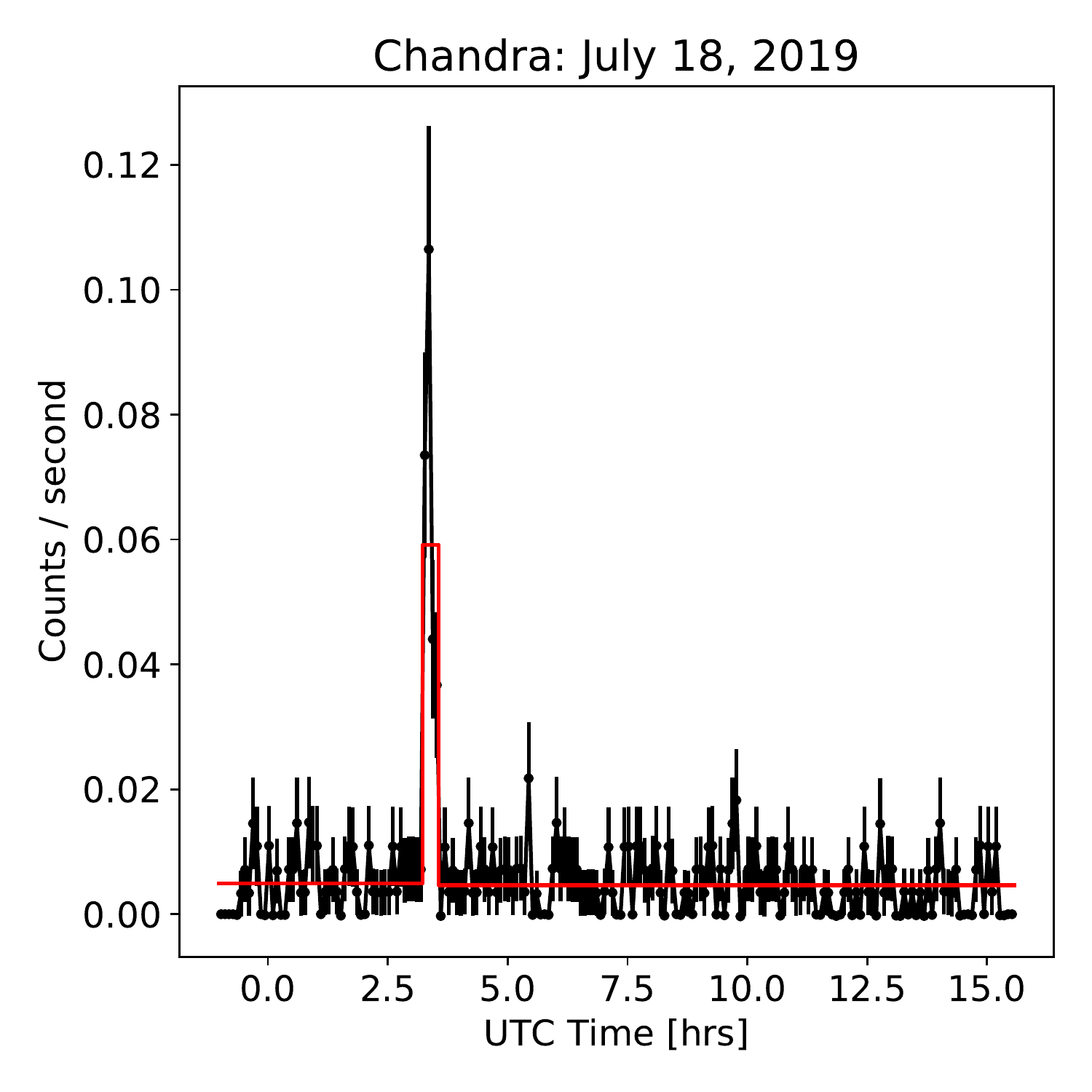}
        \caption{Multi-wavelength light curves of Sgr A* on 2019 July 18. {\it Left}: Sgr A* (blue) as observed by ALMA; the phase calibrators J1700 and J1733 (red and green, respectively) are shown. A 60-second binning time is used. {\it Center}: De-reddened \Sp\ light curve of Sgr A* at 60-second binning (black) and 5-minute binning (red). {\it Right}: \Ch\ 2-8 keV background-subtracted light curve of Sgr A* at 5-minute binning (black). The Bayesian Block results are plotted in red.}
        \label{fig:lcs}
    \end{figure*}

        To detect flaring emission, we use an open-source Bayesian blocks Python module called \texttt{bblocks}\footnote{The open-source Bayesian Blocks algorithm produced by Peter Williams is located here: \url{https://pwkit.readthedocs.io/en/latest/science/pwkit-bblocks/}.}  \citep{Scargle1998, Scargle2013, BBlocks} with a false positive rate of $p_0=0.05$. The light curves of Sgr A* and the Bayesian block results are shown in Figure \ref{fig:lcs} (right).
        
\section{Results}\label{sec:results}
        \begin{figure*}
            \centering
            \includegraphics[width=0.3\textwidth]{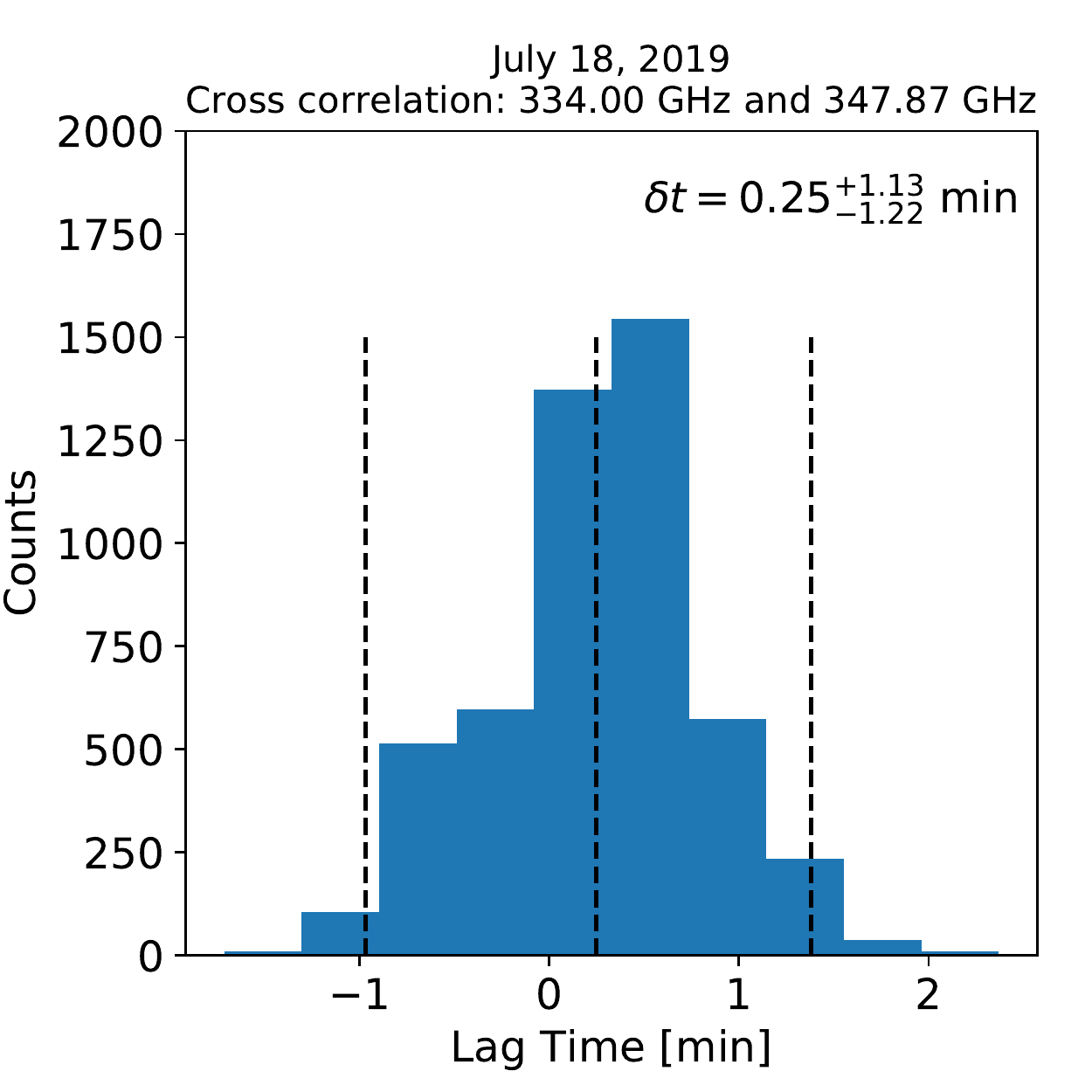}
            \includegraphics[width=0.3\textwidth]{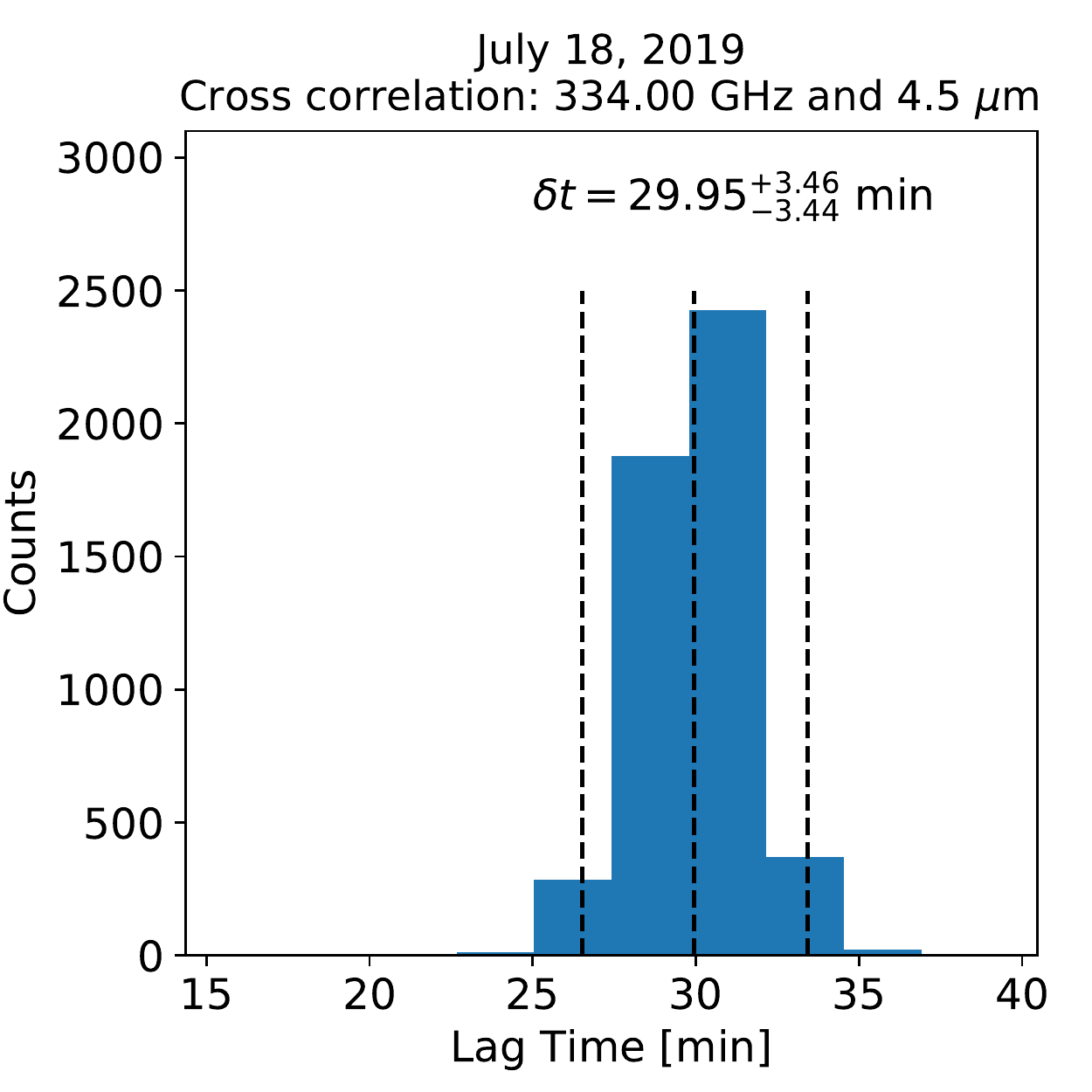}
            \includegraphics[width=0.3\textwidth]{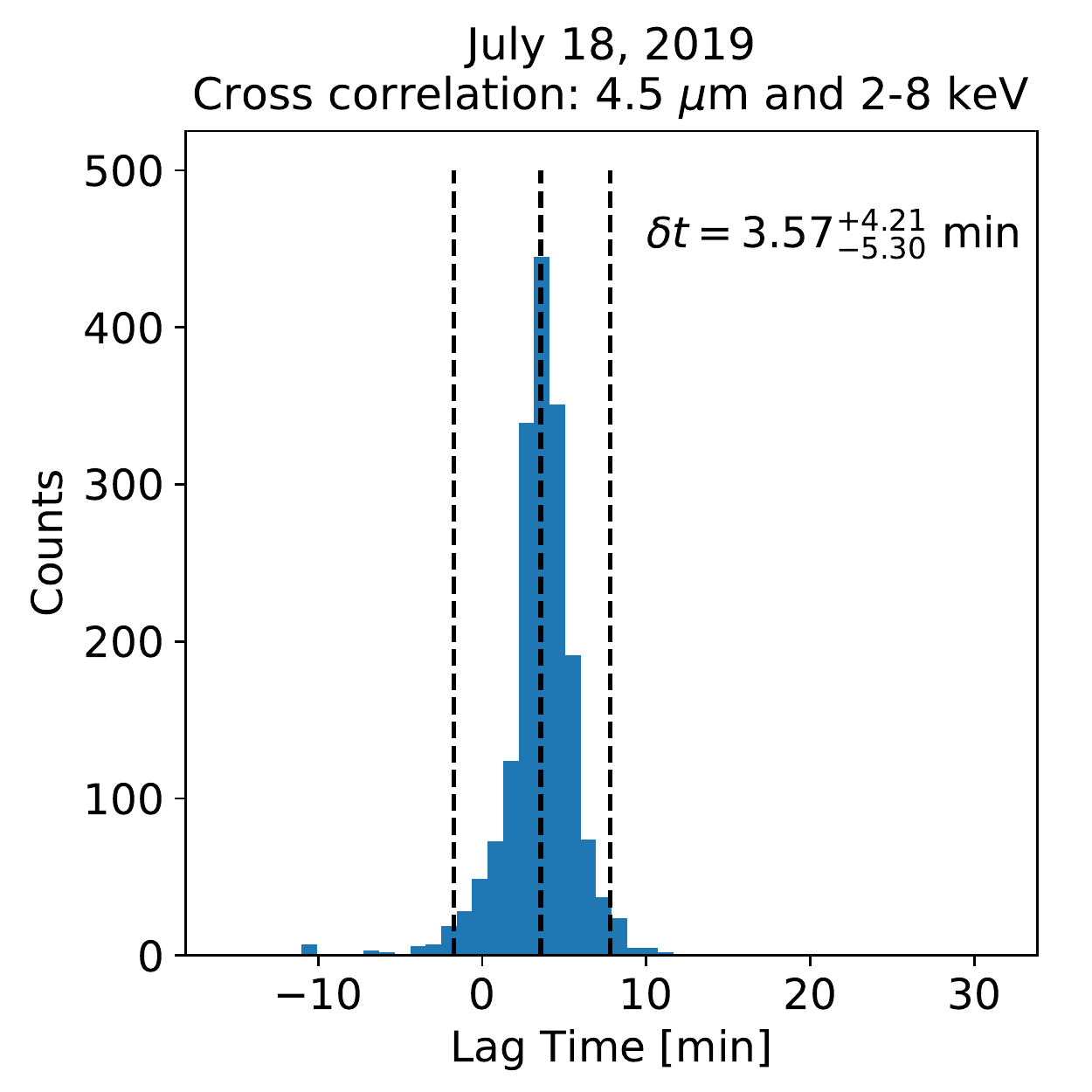}
            \caption{The centroid histograms of cross correlations between pairs of light curves using the \texttt{pyCCF} Python package \citep{PYCCF}. The $95\%$ ($2\sigma$) confidence interval (CI) is shown denoted. {\it Left}: The cross correlations between most widely separated ALMA frequencies. {\it Center}: The cross correlation histogram between the lowest ALMA frequency and \Sp\ light curves; the ALMA light curve peaks less than $30$ minutes after the \Sp\ data. {\it Right}: The cross correlation histogram between \Sp\ and \Ch, where the flaring peaks simultaneously occur.}
            \label{fig:july18_xcorr}
        \end{figure*}
        
    The ALMA, \Sp, and \Ch\ light curves for these observations all show the presence of flaring emission. Unfortunately, the submm flare peak appears to occur between ALMA execution blocks. However, a sufficient amount of the flare is present to complete cross-correlations with the other simultaneous observations. Throughout this paper, we use the \texttt{pyCCF} \citep{Peterson1998, PYCCF} Python module with 5000 Monte Carlo simulations to check for possible delays in the peak emission. A histogram of the 5000 centroid lag times is made, and we characterize the spread in the time delay using the $95\%$ ($2\sigma$) confidence interval (CI). We consider a statistically significant time lag if $0$ is not contained within this interval. The mean and 95\% CI error range is denoted on each histogram plot.
    
\begin{figure*}
            \centering
            \includegraphics[trim=0.5cm 0.6cm 0.25cm 0.5cm, clip, width=0.305\textwidth]{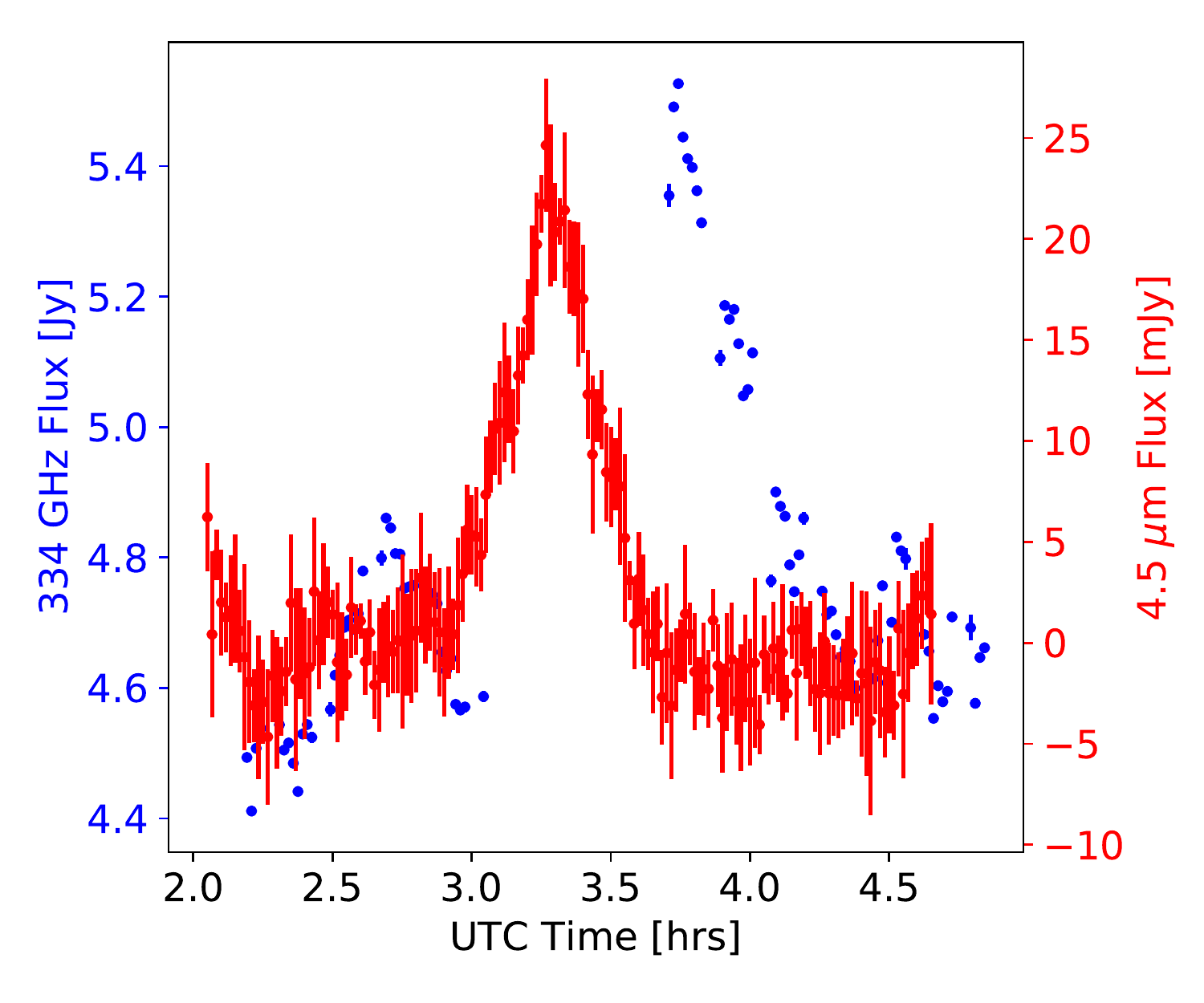}
            \includegraphics[trim=1.5cm 0.1cm 0 1.25cm, clip, width=0.53\textwidth]{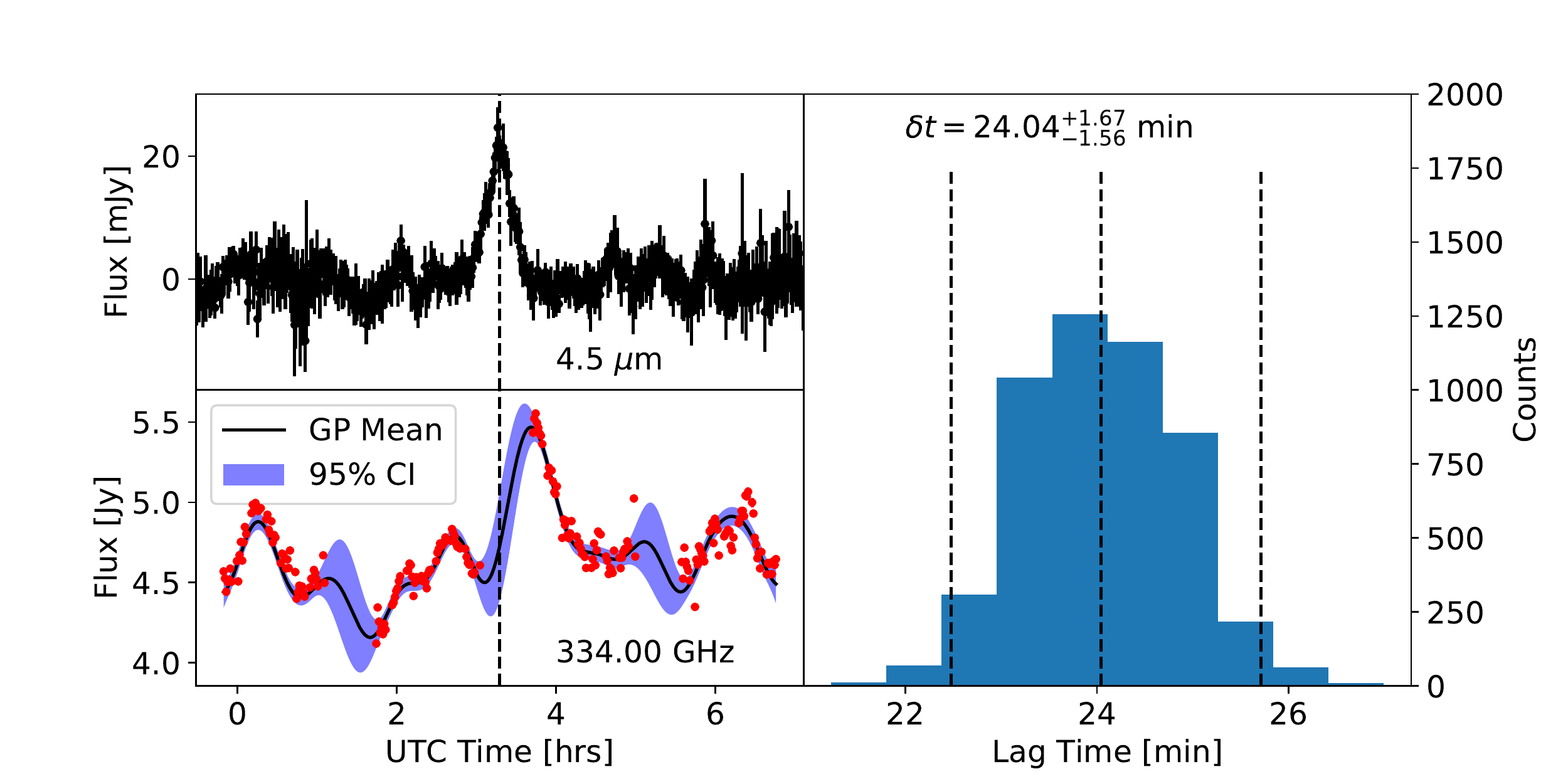}
            \caption{{\it Left}: Close-up plot of ALMA (blue) and \Sp\ (red) flaring emission detected on this night of observation. {\it Right}: The cross-correlation analysis between the \Sp\ data and Gaussian Process-regressed ALMA light curve. In the bottom left panel, the solid black line represents the mean Gaussian Process light curve, and the shaded purple region is the 95\% CI for the regressed data. The right panel shows the cross-correlation histogram between the two light curves.}
            \label{fig:july18_gp}
    \end{figure*}

    We check for, but do not detect, a time delay between the most widely separated ALMA frequencies; the histogram is shown in Figure \ref{fig:july18_xcorr} (left).
    %We do not find evidence of a time delay between the ALMA frequencies. 
    We find a time delay of $\delta t=29.95^{+3.46}_{-3.44}$ minutes between the $334$ GHz and mid-IR light curves, where the submm emission is lagging (Figure \ref{fig:july18_xcorr}, center). As a secondary check, we use the \texttt{ZDCF} cross correlation code \citep{Alexander1997} along with the \texttt{PLIKE} program \citep{Alexander2013} on the submm and mid-IR light curves. We find a time delay of $\delta t=29.51^{+6.28}_{-6.02}$ minutes, consistent with the value from \texttt{pyCCF}. A close-up plot of these two light curves is shown in Figure \ref{fig:july18_gp} (left). However, as the full submm flare was not observed, the measured time delay between the submm and mid-IR is an upper limit to the true value. In this case, the $2\sigma$ CI quoted only measures the statistical error of the observed peaks within the photometric uncertainties and does not characterize the systematic error caused by an incomplete light curve sampling.
    
    \citet{Eckart2008} detected delayed flaring emission at $345$ GHz relative to near-IR data on the order of $1.5\pm0.5$ hours. \citet{Eckart2009} lists modeled time delays between the IR to submm/radio bands, which range from $0.3$ to $1.7$ hours. %\citet{Fazio2018} detected an approximate $20$ minute time delay between \Sp\ and SMA observations at $345$ GHz. 
    The time delay upper limit we calculate between the \Sp\ and ALMA data here is well within the range of those previously published in the literature.
    
    We find that the 2-8 keV flare is simultaneous with the mid-IR data using \texttt{pyCCF} (Figure \ref{fig:july18_xcorr}, right) and agrees with the \texttt{ZDCF} analysis that we perform ($\delta t = 1.10^{+8.85}_{-4.29}$ minutes). \citet{Boyce2019} analyzed more than $100$ hours of simultaneous \Sp\ and \Ch\ data. Their analysis found the X-ray data to lead the IR flaring emission by $10-20$ minutes at the $68\%$ confidence level. However, they note the time delays of the detected flares contain $0$ at a confidence level of $99.7\%$. Our cross correlation results match with this more broad study of IR/X-ray time lags and  with the previous analysis of this data (\citealp{Gravity2021}, H. Boyce et al., in preparation). Additionally, the observed flaring emission in the Gravity K- (2.2 \micron) and H-filters (1.63 \micron) are simultaneous with the \Sp\ observations \citep{Gravity2021}. The lack of time delay at these higher frequencies matches expectations from the adiabatically expanding picture, where the IR data are optically thin.
    
    As noted above, due to non-continuous ALMA execution blocks, the true submm flare peak may not have been observed. We use a Bayesian analysis technique known as Gaussian Process Regression from the \texttt{sklearn} Python package \citep{scikit-learn} to interpolate the missing data. Our kernel is a compound radial-basis + constant function, which we use to fit the lowest frequency ALMA data. In the lower-left panel of Figure \ref{fig:july18_gp} (right), we show the observed ALMA data (red) and the mean Gaussian Process fit (black). The shaded purple region is the 95\% CI of the fit. We complete a cross-correlation between ALMA and \Sp\ using \texttt{pyCCF} with the Gaussian Process-fitted data, and their 68\% ($1\sigma$) CI errors appear to be consistent with the analysis above. The cross-correlation histogram is shown in the right panel, where we find $\delta t=24.04^{+1.67}_{-1.56}$ minutes. %This is value is similar to the time delay upper limit found using \texttt{pyCCF} and \texttt{ZDCF} above.

    Both methods find evidence for time delayed emission between the peaks of the submm and mid-IR data. Neither technique can determine if the delay is conclusive, however. The three test models presented by \citet{Gravity2021} are not able to produce time delayed emission, which is indicative of an optically thick and dynamically evolving source, nor can it match the observed peak $334$ GHz flux. Both point toward submm emission production via an optically thick adiabatically-expanding plasma \citep{VDL1966, FYZ2006b}. We consider this possibility in Section \ref{sec:analysis}.
    
\section{Analysis}\label{sec:analysis}
    In Section \ref{sec:results}, we applied two different methods to check the observed time delay between the submm and mid-IR light curves. However, the gap in our submm data near the peak allows for a range of possible conditions, namely, both simultaneous and time-delayed emission relative to the mid-IR data. If the submm emission is time delayed, the total peak flux is approximately $5.5$ Jy (requiring a peak flare flux $\approx0.9$ Jy). If the emission is simultaneous, the total flux is $\approx6$ Jy with a peak flare flux of $\approx1.4$ Jy. This latter value is estimated using a linear extrapolation of the submm data between t=3:36:00-4:12:00 hrs UTC. Note that peak flare fluxes for the simultaneous and time-delayed cases were calculated assuming a quiescent flux of $\approx4.6$ Jy, approximated from the ALMA light curve. 
    
    \subsection{Can the Gravity model reproduce the submm emission?}\label{ssec:plc}
        We focus on the parameters of the synchrotron model PLCool$\gamma_{max}$ (PLC) as described by \citet{Gravity2021}, which they note is their best-fitting model. With no knowledge of the submm emission, a synchrotron-powered flare (with a cooling break and high-energy cutoff) can describe the IR and X-ray data. However, our analysis above shows a non-zero time delay upper limit between the mid-IR and submm bands. In this case, a pure synchrotron model is insufficient to describe the submm emission.
        
        The flare detected using \Sp\ at $4.5$ \micron~had a redden-corrected flux of approximately $20$ mJy. The corresponding Gravity K- and H-filters both observed redden-corrected fluxes of nearly $15$ mJy. This is consistent with an electron power-law index of $p=2$ ($N(E)\propto E^{-p}$) and implies an IR spectrum of $S_\nu\propto\nu^{-0.5}$. The other two parameters in their model are the radius of the flaring region and magnetic field strength, for which they choose $1$ $R_\text{S}$ (Schwarzschild radius) and $30$ Gauss, respectively. For M$_{\text{BH}}$ = $4.297\times10^{6}$ M$_\odot$, $1~R_\text{S} = 1.27\times10^{12}$ cm. 
         
        We use these parameters in the context of the adiabatic expansion picture to determine the submm emission. For a spherical region emitting synchrotron radiation, the observed flux ($S_0$), radius ($R_0$), and optical depth ($\tau_0$) at frequency $\nu_0$ are related by:
            \begin{equation}
                S_0 = \dfrac{\pi R_0^2}{d^2}J_0\left(1-e^{-\tau_0}\right).
                \label{eq:synch_flux}
            \end{equation}
        $d$ is the distance to the Galactic Center, and $J_0$ is the source function of a power-law synchrotron source, which is a function of $p$, $\nu_0$, and $B$. Solving for the mid-IR optical depth yields $\tau_{\text{IR}} = 1.36\times10^{-8}$.
        In the adiabatic picture, $p$ and the critical optical depth (where the plasma blob becomes optically thin), $\tau_{\text{crit}}$, are related via:
        \begin{equation}
                e^{\tau_{\text{crit}}} - \left(\dfrac{2p}{3}+1\right)\tau_{\text{crit}}-1=0,
            \end{equation}
        \citep{FYZ2006b}. 
        %For $p=2$,. 
        The flux and opacity at any radius $R$ and frequency $\nu$ are determined by
            \begin{equation}
                S_\nu(R) = S_0\left(\dfrac{\nu}{\nu_0}\right)^{5/2}\left(\dfrac{R}{R_0}\right)^3\dfrac{1-e^{-\tau_{\nu}}}{1-e^{-\tau_0}},
            \end{equation}
        and
            \begin{equation}
                \tau_\nu(R) = \tau_{0}\left(\dfrac{\nu}{\nu_0}\right)^{-(0.5p+2)}\left(\dfrac{R}{R_0}\right)^{-(2p+3)}.
            \end{equation}
        With $\tau_{\text{IR}} = 1.36\times10^{-8}$, $\nu_0 = 66.6$ THz ($c/\nu_0=4.5$ \micron), $p=2$, $R = R_{\text{S}}$, and a critical optical depth of $\tau_{\text{crit}} = 1.51$, the submm opacity at $334$ GHz is $\tau_{\text{submm}} = 0.11$. Since $\tau_{\text{submm}} < \tau_{\text{crit}}$, the submm emission is optically thin and expansion cannot produce a time-delayed submm flare, these parameters produce a submm peak that is simultaneous with the mid-IR. At $334$ GHz, the peak flare flux is $0.27$ Jy, which underestimates the submm peak flux by a factor of $\approx2$. Therefore, we consider models with steeper spectra ($p>2$), and use the adiabatic expansion model to calculate physical parameters consistent with the submm and IR data in both cases of simultaneous and time-delayed emission.

    \subsection{Case 1: Optically Thin Submm Emission}
        \begin{figure*}
            \centering
            \includegraphics[width=0.425\textwidth]{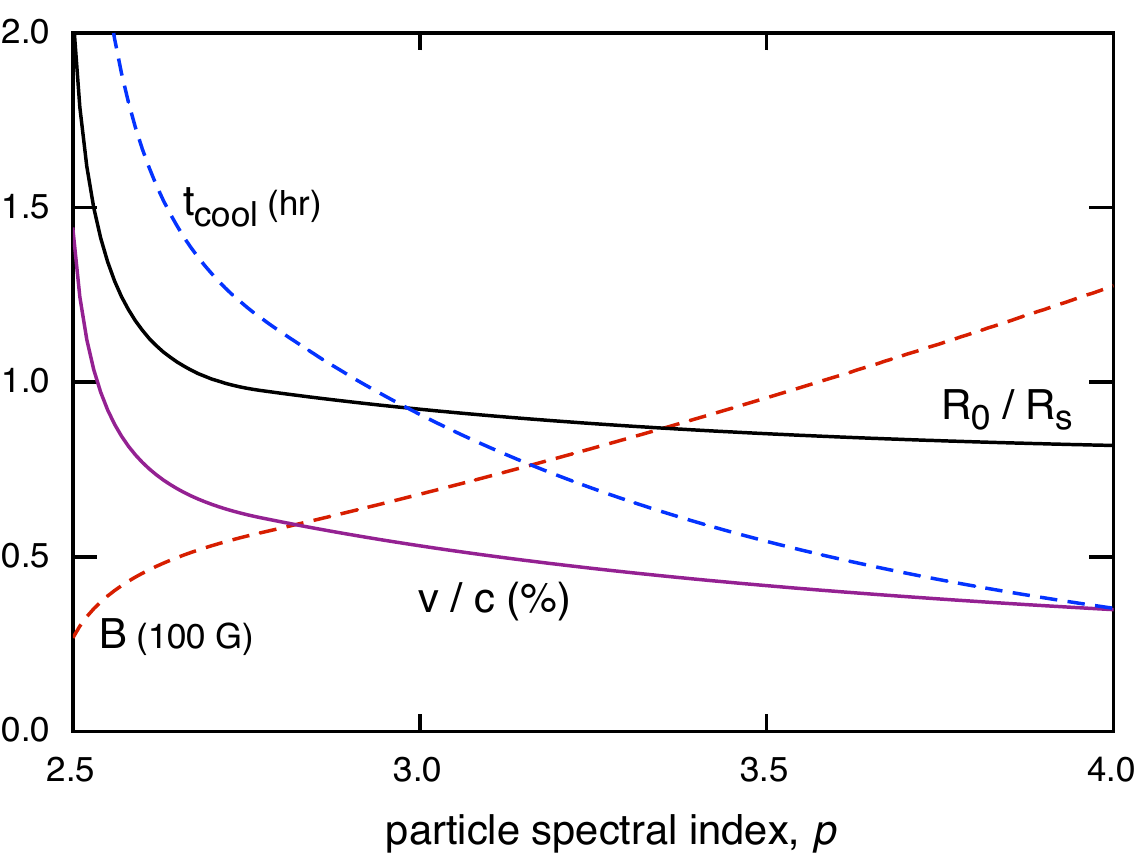}
            \includegraphics[width=0.45\textwidth]{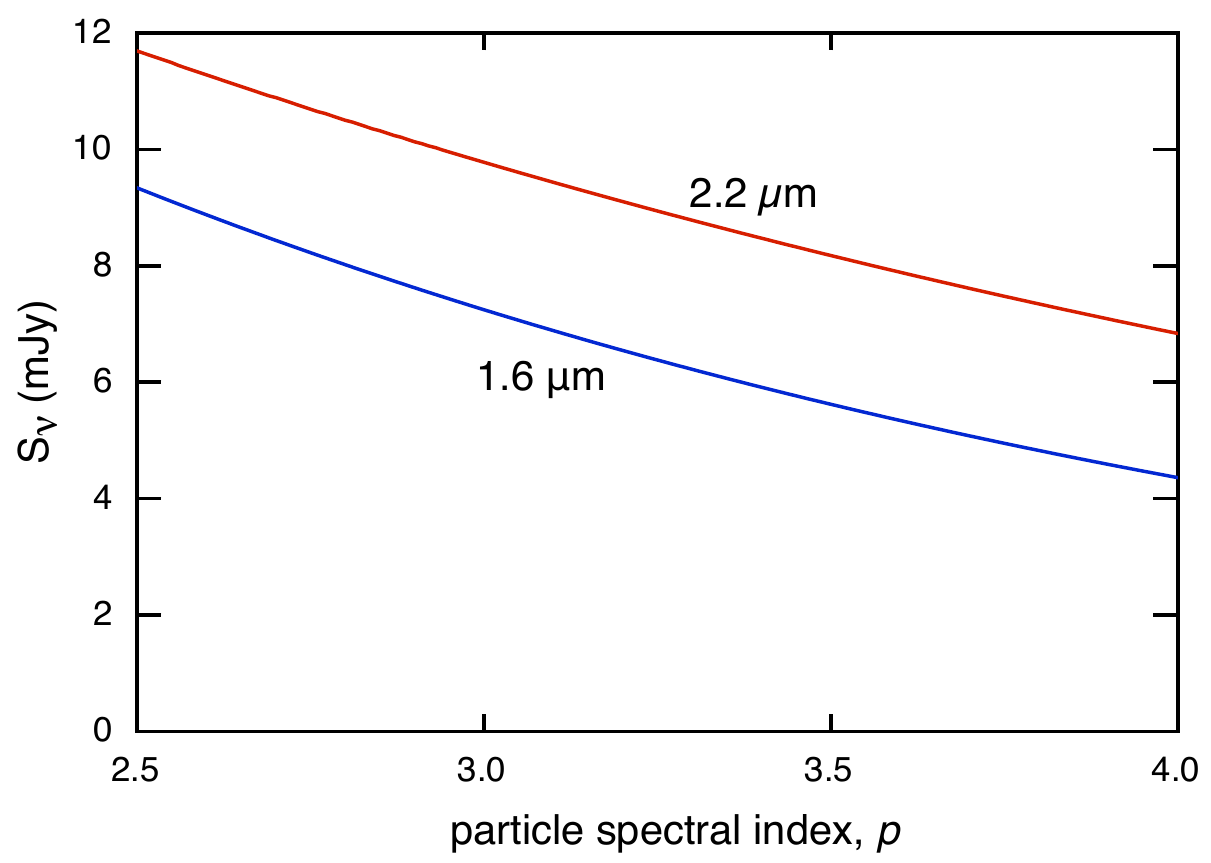}
            \caption{Predicted parameters of the flaring region assuming peak $334$ GHz and $4.5~\micron$ fluxes of $1$ Jy and $20$ mJy, respectively. {\it Left}: Physical parameters of the adiabatic expansion model as a function of $p$. The synchrotron cooling time (blue), initial radius (in terms of $R_{\text{S}}$, black), equipartition magnetic field strength (red), and expansion velocity (purple) are plotted. {\it Right}: The predicted peak fluxes for the K- ($2.2~\micron$) and H- ($1.63~\micron$) filters as a function of $p$.}
            \label{fig:physical_params}
        \end{figure*}
        \begin{figure*}
            \centering
            \plottwo{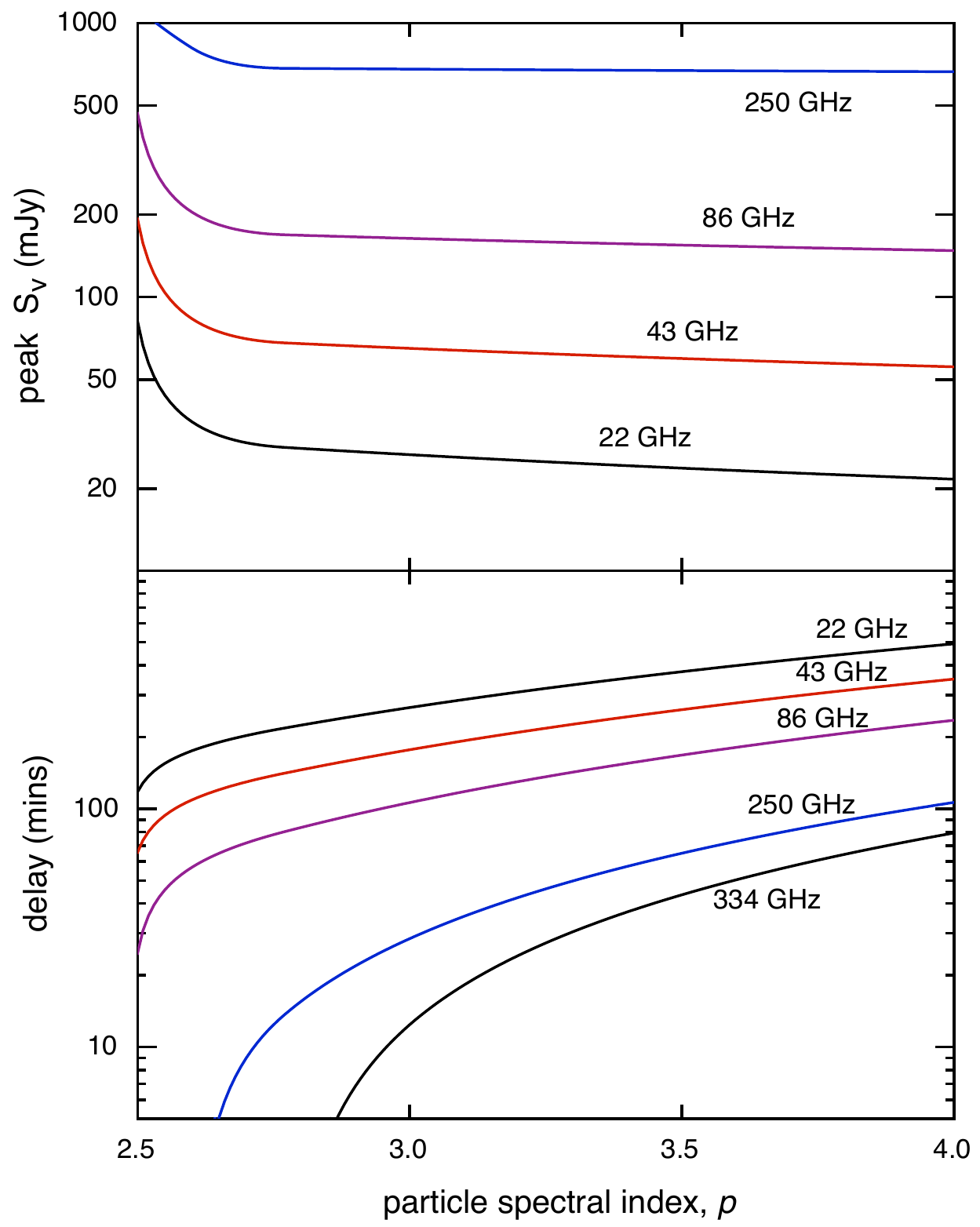}{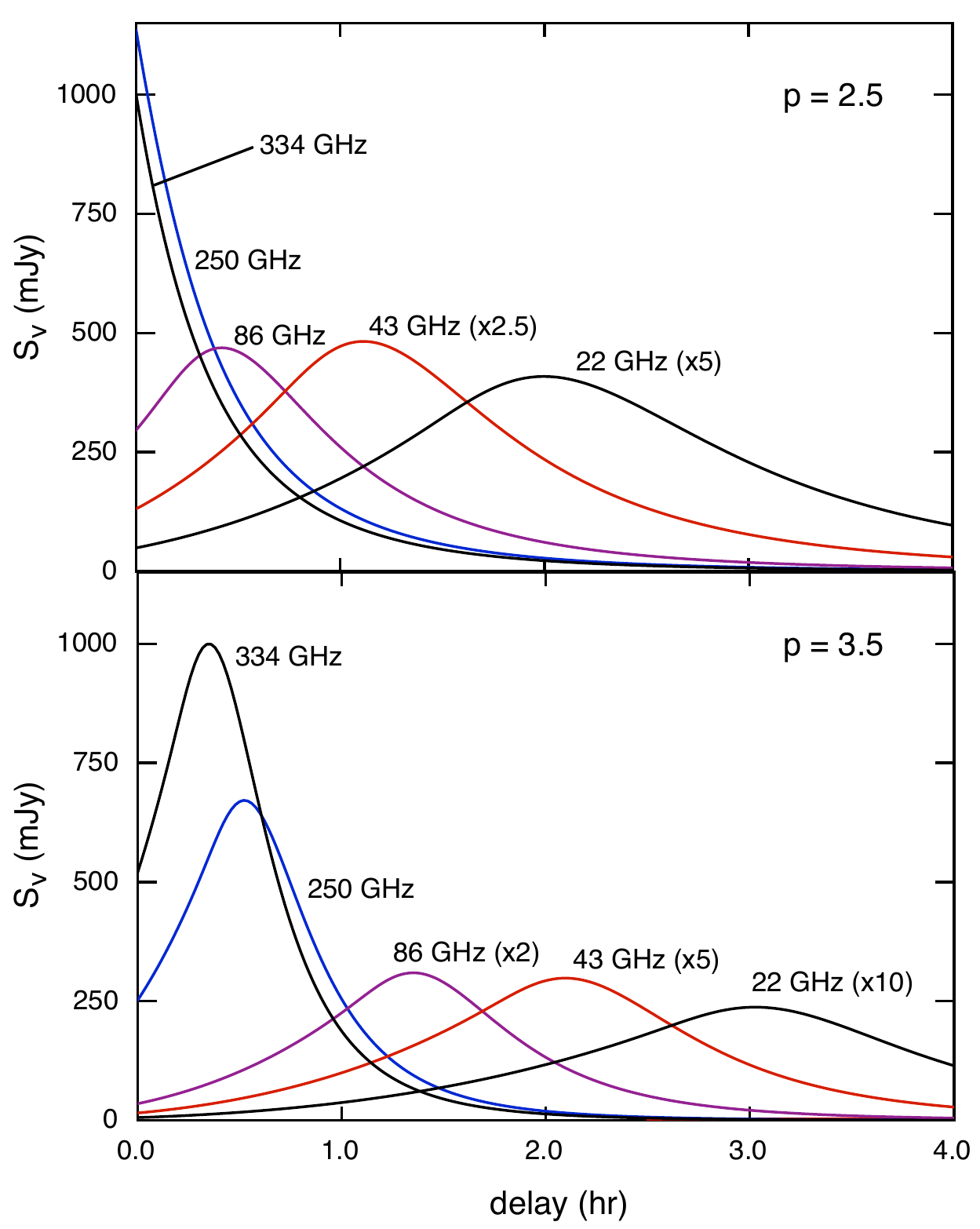}
            \caption{{\it Left}: Predicted peak fluxes (top) at four representative radio, mm, and submm wavelengths. Predicted time delays (bottom) relative to the $4.5~\micron$ emission as a function of $p$. {\it Right}: Light curves for the 4 representative frequencies and the observed submm data observed here for $p=2.5$ (top) and $p=3.5$ (bottom).}
            \label{fig:predicted_lcs}
        \end{figure*}
        The choice of $p$ will affect if the submm emission is delayed relative to the mid-IR. Additionally, it will influence physical parameters such as the equipartition magnetic field strength, cooling time, and initial radius. We show how these parameters  depend on $p$ in Figure \ref{fig:physical_params} (left). In these illustrative calculations, we assume the peak flare fluxes at $334$ GHz and $4.5~\micron$ are $1$ Jy and $20$ mJy, respectively. When calculating the equipartition magnetic field strength, $B_{\text{eq}}$, we assume the electron power-law index is valid between energies of $10$ MeV to $1$ GeV $(20 \leq \gamma_e \leq 2000)$, and neglect the presence of protons or thermal electrons. To calculate the expansion velocity and velocity-dependent parameters (such as the time delay) in Figures \ref{fig:physical_params} and \ref{fig:predicted_lcs}, we use a $15$ minute decay timescale. This timescale is estimated from the submm flare's decay rate (Figure \ref{fig:july18_gp}, left) instead of the less-certain submm/IR time delay.
        
        In the optically thin case, the flare submm-to-IR spectrum is consistent with $S_\nu\propto\nu^{-0.75}$ (using the peak flare fluxes at $4.5~\micron$ and $334$ GHz noted above). Such an optically thin spectrum is produced when $p=2.5$. For this value for $p$, the equipartition magnetic field strength is $B_{\text{eq}}\approx27$ Gauss with an initial radius of $R\approx2~R_{\text{S}}$. These parameters predict K- and H-filter peak fluxes of $\approx12$ mJy and $\approx9$ mJy, respectively (Figure \ref{fig:physical_params}, right). The K-filter flux is close to the observed value, but the H-filter flux is underestimated. This is possibly due to a systematic error in the marginal H-filter uncertainty estimate \citep{Gravity2021}. However, the spectrum may flatten between the submm and IR as it does between the IR and X-ray \citep{Gravity2021}.
        
        The synchrotron cooling time for the electrons dominating the submm emission exceeds $2$ hours, much longer than the observed flare lifetime, suggesting that the submm decay is either due to the escape of synchrotron-emitting electrons from the source region or by adiabatic cooling as the region expands. We focus on the latter scenario here as it leads to testable predictions; the former possibility is poorly constrained. A factor of $2$ decline in the flare flux occurs over approximately $15$ minutes. With $S_\nu(t)\propto R^{-2p} = R^{-5}$ in the optically thin case, we estimate an expansion velocity $v_{\text{exp}} = (2^{1/5}-1)R_0/(900\text{ seconds}) = 0.014c$ at lower frequencies, within the typical range $v_{\text{exp}} = 0.003c-0.1c$ inferred for previous flares at radio/mm frequencies \citep{FYZ2008}. This model cannot explain the $\approx30$ minute time delay upper limit between submm and IR wavelengths, but radio/mm observations would be delayed (Figure \ref{fig:predicted_lcs}, top right panel).
        
    \subsection{Case 2: Time-Delayed Submm Emission}
        For $p\geq2.8$, $\tau_{\text{submm}}$ at the time of the $4.5~\micron$ peak is high enough to delay the submm peak due to the transition from an optically thick to thin plasma as the synchrotron source expands. For values of $p$ that are just above $2.8$, the expansion speed is too slow to match the $30$ minute lag upper limit of the $334$ GHz light curve relative to the IR (lower left panel of Figure \ref{fig:predicted_lcs}). A steeper electron spectrum with $p\approx3.5$ is preferred to match both the $30$ minute time delay upper limit and the $15$ minute decay timescale. Then, $R\approx0.8 R_\text{S}$, $B_{\text{eq}}\approx90$ Gauss, and $v_{\text{exp}}\approx0.004c$. The light curves for the representative radio/mm/submm frequencies are shown in Figure \ref{fig:predicted_lcs} (bottom right).

        Simultaneous, multi-wavelength observations of Sgr A* at all accessible frequencies should not be undervalued. Figure \ref{fig:predicted_lcs} shows that the ambiguity of the models for the submm-IR time delay can be broken by observing at least one lower frequency.
        
\section{Summary and Conclusions}\label{sec:conclusions}
    We have reported and analyzed the submm light curve of Sgr A* observed as part of a global multi-wavelength campaign on 2019 July 18. We confirm that the mid-IR data, observed with the Spitzer Space Telescope, is simultaneous with soft X-ray emission detected with the Chandra X-ray Observatory. Our analysis finds a time delay between the submm light curve, taken with ALMA, and the \Sp\ mid-IR light curve. However, the submm data do not sample the entire flare, so the measured time delay is an upper limit to the true value. We analyze the submm and mid-IR emission using adiabatically expanding synchrotron plasma models. We prefer an electron energy spectrum with $p=2.5$ if the submm emission is initially optically thin. We do not make conclusions about the electron spectrum at high energies, as a cooling break and high energy cut-off is preferred \citep{Gravity2021} but is not modeled here. Since the submm and mid-IR time delay is not certain, we calculate physical parameters of the flaring emission for different values of the electron power-law index, $p$, which cover both optically thick and thin models. If the submm-to-IR time delay is truly absent, an adiabatically-expanding plasma with $p=2.5$, $R\approx2~R_\text{S}$, equipartition magnetic field strength of $B_{\text{eq}}\approx27$ Gauss, and expansion speed $v_{\text{exp}}\approx0.014c$ can model the data. 
    If the submm emission is delayed by 30 minutes relative to the mid-IR, an adiabatically-expanding plasma with initial radius $R\approx0.8~R_{\text{S}}$, $p=3.5$, $B_{\text{eq}}\approx90$ Gauss, and $v_{\text{exp}}\approx0.004c$ is preferred. In either case, the need for an expanding source region is clear. We showed that multi-wavelength observations using at least one lower frequency band would have been able to break the degeneracy and should be part of future multi-wavelength campaigns.

\begin{acknowledgements}
The authors are greatly appreciative to the \Sp\ Sgr A* Collaboration for allowing us to use their data in this analysis. The authors are particularly indebted to Joe Hora for answering our questions regarding the Spitzer data. We would also like to thank the anonymous referee for their very helpful and constructive comments. This work is partially supported by the grant AST-0807400 from the
the National Science Foundation. J.M. gratefully acknowledges support for this work, which was provided by the NSF through award SOSP21A-003 from the NRAO. The National Radio Astronomy Observatory is a facility of the National Science Foundation operated under cooperative agreement by Associated Universities, Inc. This paper makes use of the following ALMA data: ADS/JAO.ALMA\#2018.A.00050.T. ALMA is a partnership of ESO (representing its member states), NSF (USA) and NINS (Japan), together with NRC (Canada), MOST and ASIAA (Taiwan), and KASI (Republic of Korea), in cooperation with the Republic of Chile. The Joint ALMA Observatory is operated by ESO, AUI/NRAO and NAOJ. This work is based on archival data obtained with the Spitzer Space Telescope, which was operated by the Jet Propulsion Laboratory, California Institute of Technology under a contract with NASA. The scientific results reported in this article are based in part on data obtained from the Chandra Data Archive. This research has made use of software provided by the Chandra X-ray Center (CXC) in the application packages CIAO. This research has made use of data obtained from the Chandra Source Catalog, provided by the Chandra X-ray Center (CXC) as part of the Chandra Data Archive. This research was supported in part through the computational resources and staff contributions provided for the Quest high performance computing facility at Northwestern University which is jointly supported by the Office of the Provost, the Office for Research, and Northwestern University Information Technology.
\end{acknowledgements}
\software{CASA \citep{McMullin2007}, CIAO, bblocks \citep{Scargle1998, Scargle2013, BBlocks}, pyCCF \citep{Peterson1998, PYCCF}, ZDCF \citep{Alexander1997}, PLIKE \citep{Alexander2013}, sklearn \citep{scikit-learn}, Scipy \citep{SciPy2020}, Jupyter Notebook \citep{Kluyver2016}, Pandas \citep{Mckinney2010}, Matplotlib \citep{Hunter2007}}
\facilities{ALMA, Spitzer (IRAC), Chandra (ACIS)}

\bibliography{sample631}{}
\bibliographystyle{aasjournal}



\end{document}